\documentclass[aps,prd,12pt,superscriptaddress,amsfonts,amssymb,amsmath,byrevtex,showpacs,nofootinbib]{revtex4-1}
\usepackage{graphicx}
\usepackage{amssymb,amsmath,amsfonts,palatino,amsthm}
\usepackage{amssymb}
\usepackage{epstopdf}
\usepackage{color}

\begin{document}

\title{Spin-Field Correspondence}

\author{Jakub Mielczarek  \\
Institute of Physics, Jagiellonian University, {\L}ojasiewicza 11, 30-348 Cracow, Poland }

\email{jakub.mielczarek@uj.edu.pl} 

\begin{abstract}
In the recent article Phys.\ Lett.\ B {\bf 759} (2016) 424 a new class of field theories called Nonlinear Field 
Space Theory has been proposed. In this approach, the standard field theories are considered as linear 
approximations to some more general theories characterized by nonlinear field phase spaces. The case 
of spherical geometry is especially interesting due to its relation with the spin physics. Here, we explore 
this possibility showing that classical scalar field theory with such a field space can be viewed as a 
perturbation of a continuous spin system. In this picture, the spin precession and the scalar field excitations 
are dual descriptions of the same physics. The duality is studied on the example of the Heisenberg model. 
It is shown that the Heisenberg model coupled to a magnetic field leads to a non-relativistic scalar field 
theory, characterized by quadratic dispersion relation. Finally, on the basis of analysis of the relation 
between the spin phase space and the scalar field theory we propose the \emph{Spin-Field correspondence} 
between the known types of fields and the corresponding spin systems. 
\end{abstract}

\maketitle 

\section{Introduction}

The phase space of a classical spin is a two-sphere, $S^2$ (see e.g. \cite{Kirillov}). Because the phase space is a 
symplectic manifold it has to be equipped with the closed symplectic form $\omega$, for which the natural choice 
is the area two-form. Using the standard spherical coordinates $(\phi,\theta)$ the symplectic form can be written 
as $\omega = J \sin\theta\, d\phi \wedge d\theta$. The $J$ is a constant introduced due to dimensional reasons, 
equal to the absolute value of the spin (angular momentum). Except the poles $\theta=0,\pi$, the symplectic form $\omega$ 
is well defined and invertible, allowing for determination of the Poisson tensor $\mathcal{P}^{ij} = (\omega^{-1})^{ij}$, 
and then we can define the Poisson bracket $\{f,g\} = \mathcal{P}^{ij}(\partial_i f)(\partial_j g)$.  The Hamilton equation 
can then be defined as $\frac{d}{dt} f = \{f,H\}$, where $f$ is some phase space function and $H$ is the Hamiltonian. 

It turns out that instead of using the spherical coordinates $(\phi,\theta)$, it is useful to apply another local 
parametrization of sphere. Let us namely consider the following change of coordinates:
\begin{eqnarray}
\phi &=& \frac{q}{R_1}  \in (-\pi, \pi], \\
\theta &=& \frac{\pi}{2}-\frac{p}{R_2}  \in (0,\pi), 
\end{eqnarray}
where $q$ and $p$ are our new phase space variables and $R_1$ and $R_2$ are constants introduced 
due to dimensional reasons. Using the new variables, the symplectic form can be written as $\omega 
= \cos(p/R_2) dp \wedge d q$, where we fixed $R_1 R_2 = J$. The last condition guarantees that for 
small values of $p$ ($p \ll R_2$) the symplectic form reduces to Darboux form $\omega = dp \wedge dq$. 
Employing the new variables, the Poisson bracket related to $\omega$ can now be written as:
\begin{equation}
\{f,g\} = \frac{1}{\cos(p/R_2)}\left( \frac{\partial f}{\partial q} \frac{\partial g}{\partial p}-\frac{\partial f}{\partial p} \frac{\partial g}{\partial q}\right). 
\label{Poisson}
\end{equation} 

The $(q,p)$ variables, similarly as the $(\phi,\theta)$ pair, are defined only locally. It is, however, often convenient 
to work with the globally defined functions. The most important example of such functions are components of the 
angular momentum vector $\vec{J}=(J_x,J_y,J_z)$, which can be defined as follows: 
\begin{eqnarray}
J_x &:=& J \sin\theta \cos\phi = J \cos \left( \frac{p}{R_2} \right) \cos \left( \frac{q}{R_1} \right) =J\left(1-\frac{p^2}{2R_2^2}-\frac{q^2}{2R_1^2}+\mathcal{O}(4) \right) \,, \label{Jx} \\
J_y &:=& J \sin\theta \sin\phi = J \cos \left( \frac{p}{R_2} \right) \sin \left( \frac{q}{R_1} \right) = J \left( \frac{q}{R_1}+\mathcal{O}(3)\right) \,, \label{Jy}\\
J_z &:=& J \cos\theta = J \sin \left( \frac{p}{R_2} \right)=J \left( \frac{p}{R_2}+\mathcal{O}(3)\right) \,, \label{Jz}
\end{eqnarray}
together with the condition $J_x^2+J_y^2+J_z^2=J^2=const$. Here, for the purpose of our further analysis the leading 
order expansions have been included. In the atomic physics, magnetic moment couples to an external magnetic field $\vec{B}$ 
via the vector $\vec{J}$. In such a case, the Hamiltonian of interaction is 
\begin{equation}
H = - \frac{\mu}{J} \vec{J}\cdot \vec{B},
\label{HSJ}
\end{equation}
where $\mu$ is the value of the magnetic moment, which can be both positive and negative. 
Basing on (\ref{HSJ}), together with the fact that the $J_i$ components satisfy the $su(2)$ 
algebra bracket $\{J_i,J_j\}=\epsilon_{ijk} J_k$, one can write Hamilton equation for the vector $\vec{J}$ 
in the following form: 
\begin{equation}
\dot{\vec{J}} = \{ \vec{J},H\}= - \frac{\mu}{J}  \vec{B} \times \vec{J}.
\label{prec}
\end{equation}

\begin{figure}[ht!]
\centering
\includegraphics[width=15cm,angle=0]{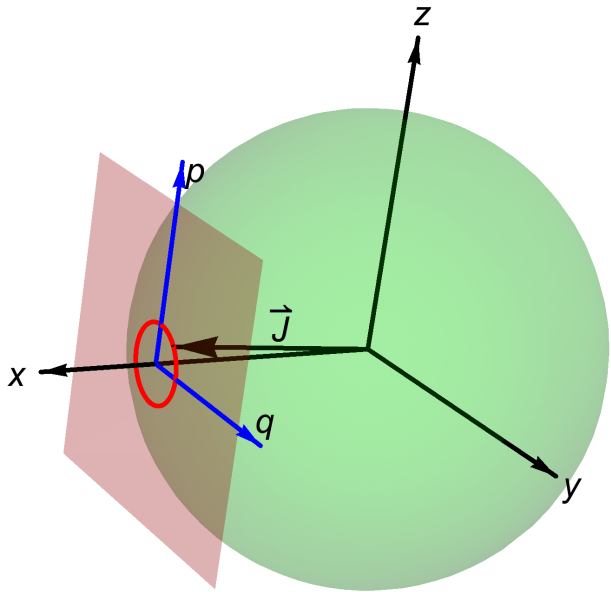}
\caption{Graphical representation of precession of the vector $\vec{J}$ around the $x$ axis
(the magnetic field $\vec{B}=(B_x,0,0)$). For small precession angles the arrow of the vector outlines a 
circle on the $(q,p)$ plane. The picture captures the idea of local approximation of the spin phase 
space ($S^2$) by the $\mathbb{R}^2$ phase space, where the relation \emph{precession of $\vec{J}$ = 
oscillation of $q$} is satisfied.} 
\label{SF}
\end{figure}

Equation (\ref{prec}) describes the precession of magnetic moment in a constant
magnetic field.  The process plays a crucial role in our construction. Firstly, let us notice 
that the arrow of the $\vec{J}$ vector goes around a circle (see Fig. \ref{SF}). Secondly, 
suppose that the precession angle (angle between the $\vec{J}$ and $\vec{B}$ vectors) is 
small. Then, let us make use of the new variables $(q,p)$ and direct the vector $\vec{B}$ 
through the origin of the $(q,p)$ coordinate system (i.e. along the $x$ axis). 

One can now easily conclude that precession of the vector  $\vec{J}$ corresponds to a circle in
the $(q,p)$ phase space. The precession (for small precession angles) is, therefore, described by 
harmonic oscillator. In order to see this explicitly, let us fix  $\vec{B}=(B_x,0,0)$ so that precession takes 
place around the origin of the $(q,p)$ coordinate system. Then, for small spin displacements from the 
equilibrium point, the Hamiltonian (\ref{HSJ}) can be written as 
\begin{equation}
H \approx  -\mu B_x \left(1-\frac{p^2}{2R_2^2}-\frac{q^2}{2R_1^2} \right) = \frac{p^2}{2m}+\frac{\omega^2q^2}{2}  + const,   
\end{equation}
where we defined the constants such that $m:= \frac{R^2_2}{\mu B_x}$  and $\omega:= 
\frac{\sqrt{\mu B_x}}{R_1}$. The constant energy trajectory is in general an ellipse in the $(q,p)$ variables 
or a circle if the variables are appropriately rescaled, namely if the pair $(p/\sqrt{m}, \omega q)$ 
is considered.

The picture can be now generalized to the area of field theory. For this purpose, let us consider a 
continuous spin distribution, which may be viewed as an approximated description of some 
discrete spin systems studied experimentally. Then, at any space point the following identification 
can be performed:  
\begin{eqnarray}
&&q\rightarrow \varphi({\bf x},t),  \label{qphi} \\
&&p\rightarrow \pi_\varphi({\bf x},t). \label{ppi}
\end{eqnarray}
The continuous spin system is, therefore, in correspondence with the scalar field theory 
with the spherical field phase space. This is basically because dimension of the scalar field 
phase space $\Gamma^{\varphi}_{\bf{x}}$ at any point is equal to the dimension of the spin 
phase space:
\begin{equation}
\text{dim}(\Gamma^{\varphi}_{x})= \text{dim} S^2 .
\end{equation}

Depending on the particular form of the interactions between the spins, different types of the field 
theories with the bounded field spaces can be reconstructed. In what follows, we consider 
the Heisenberg model of the spin system and find what is the field theoretical counterpart in such a case.   

\section{The Heisenberg model}

The Heisenberg model with an external magnetic field $\vec{B}$ can be defined by 
the following Hamiltonian:
\begin{equation}
H =-c_1 \sum_{i,j} \vec{J}_i \cdot  \vec{J}_j -c_2 \sum_i \vec{B} \cdot  \vec{J}_i, 
\label{Heisenberg}
\end{equation}
where $c_1$ and $c_2$ are real-valued coupling constants and the first sum is 
performed over the closest neighbors. The indices $i,j$ are labeling different spins (not the 
spin vector components) distributed on regular three-dimensional lattice. In the physical picture of 
ferromagnets the spins $\vec{J}_i$ are accompanied by magnetic moments. 

The discrete spin system described by the Heisenberg model (\ref{Heisenberg}) can be 
approximated by continuous spin distribution. In such a limit, the spin labels $i$ are replaced 
by the position vector $\vec{x}$ and neglecting the higher-order derivatives the corresponding continuous 
version of the Hamiltonian (\ref{Heisenberg}) can be written as:
\begin{equation}
H = \int d^3x \mathcal{H} =  \int d^3x \left[ \tilde{c}_1 (\nabla \vec{J} )^2 - \tilde{c}_2  \vec{B} \cdot  \vec{J}  \right],
\label{HeisenbergCont}
\end{equation}
together with the condition $\vec{J}\cdot \vec{J} =J^2 = const$. The sign change of the first factor with respect to 
expression (\ref{Heisenberg}) appears a consequence of the integration by parts. Here, $\tilde{c}_1$ and $\tilde{c}_2$ are 
new coupling constants determined by $c_1$, $c_2$ and a lattice spacing parameter. 

Let us now chose direction of the $\vec{B}$ vector along the $x$ axis. In this way, the set of field variables (\ref{qphi},\ref{ppi}) 
discussed in the previous section can be introduced. Additionally, it is worth mentioning here that expression
(\ref{HeisenbergCont}) interpreted not as a Hamiltonian but as a Lagrangian gives us a certain non-linear sigma 
model \cite{GellMann:1960np}.    

Using relations (\ref{Jx}-\ref{Jy}) together with (\ref{qphi}) and (\ref{ppi}), 
we find that the scalar field Hamiltonian density resulting from (\ref{HeisenbergCont}) is up 
to the quadratic terms: 
\begin{equation}
\mathcal{H}  = \frac{\pi_\varphi^2}{2}+\frac{1}{2}(\nabla  \varphi)^2+\frac{1}{2}M^2 \varphi^2 +\frac{1}{2 M^2}(\nabla  \pi_\varphi)^2,
\label{HamField}
\end{equation}
where we fixed the first three terms to have the standard form. This leads to the following relations 
between the parameters:
\begin{eqnarray}
M &:=& \tilde{c}_2 B_x, \\
R_1 &=& \sqrt{\frac{J}{M}}, \\ 
R_2 &=& \sqrt{JM}, \\
\tilde{c}_1 &=& \frac{1}{2JM},
\end{eqnarray} 
where we also used the condition $J=R_1R_2$ coming from the small field values limit of the symplectic form. 
Substantially, $J$ and $M$ can be treated as two independent parameters. However, then we have
to keep in mind that the coupling constants $\tilde{c}_1$ and  $\tilde{c}_2$ may vary as a
functions of $J$ and $M$. On the other hand one may assume that $\tilde{c}_1$ is fixed,  
and then $\tilde{c}_2 = \frac{1}{2 B_x \tilde{c}_1 J}$. Actually, the presence of $J$ in the 
expression for $\tilde{c}_2$ is a good sign. The $J$ factor in the denominator compensates 
the $J$ coming from vector $\vec{J}$ in the interaction term with the magnetic field. This is in 
agreement with the experimental evidence that the interaction energy does not depend on 
the absolute value of $\vec{J}$.   

The novelty in the Hamiltonian (\ref{HamField}) is the presence of the factor 
$\frac{1}{2 M^2}(\nabla  \pi_\varphi)^2$, which is in some sense dual to the 
kinetic term $\frac{1}{2}(\nabla  \varphi)^2$. The term $\frac{1}{2 M^2}(\nabla  \pi_\varphi)^2$ is, 
however, explicitly breaking the Lorentz symmetry and, therefore, is not present in the standard 
field theory. It is tempting to consider the large mass limit ($M\rightarrow \infty$),
which would suppress the Lorentz symmetry breaking term. However, one has 
to keep in mind that the mass term $\frac{1}{2}M^2 \varphi^2$ becomes dominant then. 
In the case of ferromagnetic system this limit would be associated with domination of the 
coupling to external magnetic field, e.g. as a consequence of increase of the magnetic field 
strength. Whether or not the limit is relevant in the context of the relativistic limit 
will become clear while studying equations of motion and resulting dispersion 
relation for the plane waves. 

In order to derive equations of motions resulting from the Hamiltonian  (\ref{HamField})  
we have to firstly write the Poisson bracket:
\begin{equation}
\{f({\bf x}) ,g({\bf y}) \} = \int \frac{d^3{\bf z} }{\cos(\pi_{\varphi}({\bf z}) /R_2)}
\left( \frac{\delta f({\bf x})}{\delta \varphi({\bf z})} \frac{\delta g({\bf y})}{\delta 
\pi_{\varphi}({\bf z}) }-\frac{\delta f({\bf x})}{\delta \pi_{\varphi}({\bf z}) } \frac{\delta g({\bf y})}{\delta \varphi({\bf z}) }\right), 
\label{PoissonField}
\end{equation} 
which is a field theoretical version of the Poisson bracket (\ref{Poisson}), studied in the previous section.  
In the leading order, the Hamilton equations for the canonical variables are:
\begin{eqnarray}
\dot{\varphi} &=& \pi_{\varphi} -\frac{1}{M^2} \Delta \pi_{\varphi}, \label{HE1} \\ 
\dot{\pi}_{\varphi} &=& - M^2 \varphi + \Delta \varphi, \label{HE2}
\end{eqnarray}
which when combined lead to the following equation of motion for the field $\varphi$:
\begin{equation}
\ddot{\varphi}-2\Delta \varphi+M^2\varphi+\frac{1}{M^2}\Delta^2\varphi =0. 
\label{KGMod}  
\end{equation}
There are two differences between the equation (\ref{KGMod}) and the standard relativistic Klein-Gordon 
equation 
\begin{equation}
\ddot{\varphi}-\Delta \varphi+M^2\varphi=0.
\label{KGEQ}   
\end{equation}
First, is the factor $2$ in front of the Laplace operator. Second, is the Lorentz symmetry breaking term 
$\frac{1}{M^2}\Delta^2\varphi$. 

To address the issue of departure of the equation (\ref{KGMod}) from the Lorentz invariance in more 
details, let us perform the Fourier transform 
\begin{equation}
\varphi(t,{\bf x} )= \int \frac{d^3 k d\omega }{(2\pi)^4} \varphi(\omega,{\bf k} )e^{i({\bf k \cdot x}-\omega t)},    
\end{equation}
which leads to the following dispersion relation of the plane waves:
\begin{equation}
\omega^2 = 2k^2+M^2 +\frac{k^4}{M^2}.
\end{equation}
The dispersion relation can be factorized and squared leading to the following simple expression for $\omega$: 
\begin{equation}
\omega = M+\frac{k^2}{M}.
\end{equation}
The obtained quadratic dispersion relation is in agreement with the dispersion relation of magnons (spin waves) 
expected for ferromagnets \cite{SW}. Correspondingly, the group velocity of the plane waves is a linear 
function of $k$: $v_{\text{gr}} := \frac{\partial \omega}{\partial k} = 2\frac{k}{M}$. As one can easily conclude, 
the large mass limit ($M\rightarrow \infty$) has no significance in the context of relativistic limit of the theory. 
However, let us mention that model of spin system which leads to relativistic scalar field theory 
in the leading order, can be constructed \cite{SFNEW}.

All of the results presented in this section are classical. The units are such that $c=1=\hbar$.
However, one has to keep in mind that in the adopted convention the value of $J$ parameter 
is measured in the units of $\hbar$, as a result of the quantum normalization of the variables. 
The full quantum theory is not discussed here, but can be introduced using standard canonical 
methods. Worth stressing is the observation that relation between the spin ($su(2)$) and bosonic 
quantum representations is conceptually similar to the Holstein-Primakoff transformation \cite{HP} . 

\section{The Spin-Field Correspondence}

Analysis of the last two sections has shown that there is a direct relation between the 
spin system and the scalar field theory. Basically, the relation is possible because the  
dimension of the phase space of spin (angular momentum) and the dimension of the
scalar field phase space at a point are equal:
\begin{equation}
\text{dim} S^2 = \text{dim}(\Gamma^{\varphi}_{x}).
\end{equation}
Taking this into account, the angular momentum vector $\vec{J}$ of a fixed length 
$(\vec{J}\cdot\vec{J}=const)$ can be parametrized by the field variables: 
\begin{equation}
\vec{J}  \Longleftrightarrow (\varphi, \pi_{\varphi})
\end{equation}
and field theoretical Hamiltonian can be reconstructed for a given spin system model. 
As an example we have shown that the Heisenberg model coupled to a constant magnetic field 
is in correspondence with the following scalar field theory:  
\begin{equation}
H =-c_1 \sum_{i,j} \vec{J}_i \cdot  \vec{J}_{j} -c_2 \sum_i \vec{B} \cdot  \vec{J}_i  \  \Longleftrightarrow \ \mathcal{H}  
= \frac{\pi_\varphi^2}{2}+\frac{1}{2}(\nabla  \varphi)^2+\frac{1}{2}M^2 \varphi^2 +\frac{1}{2M^2}(\nabla  \pi_\varphi)^2.
\end{equation}

The question is now, if the construction can be generalized to the different types
of (bosonic and fermonic) fields, so that the relation: 
\begin{equation}
\text{Spin system Hamiltonian} \Longleftrightarrow \text{Field theory Hamiltonian} 
\end{equation}
is not reserved only to the case of the scalar field. 

Indeed, we conjecture that such a correspondence can be postulated, which will require
introduction of multiple spins per space point or for elementary cell in the discrete solid 
state model.  E.g. a spinor field, can possibly be constructed out of two scalar fields. Two different 
spins per point are, therefore, required in the model of the spin system. The same 
is expected in the case of the scalar field doublet. However, the corresponding Hamiltonian 
must then have a fundamentally different structure. In general, in order to construct 
a massive field with the angular momentum $s$ we need 
\begin{equation}
N_x = \text{dim}(\Gamma_{x})/2 = (2s+1)
\end{equation}
different spins at each point in the dual spin system. In Table \ref{tab1} we summarize 
the corresponding number of spins at each point for the five most commonly considered types of physical fields.  
\begin{table}[h!]
\caption{The correspondence between the different multiplicity of spins and the corresponding field theories.}
\centering
\begin{tabular}{|c|c|c|c|}
\hline
\textbf{Field type}	& \textbf{Field spin} (s) 	&  $\text{dim}(\Gamma_{x})$  & $N_x$ \\
\hline
Scalar field 		       & 0	 & 2  &  1\\
Spinor field		       & 1/2    & 4 &   2 \\
Vector field		       & 1       & 6 & 3\\
Rarita-Schwinger field      & 3/2     & 8  &  4\\
Tensor field                       &  2        & 10 & 5  \\
\hline
\end{tabular}
\label{tab1}
\end{table}

Of course, in the case of massless fields, some of the degrees of freedom have to be
suppressed. This can be realized either as an appropriate large spin limit of the 
corresponding field theories or assumed from the very beginning by restricting the 
number of physical degrees of freedom. In this second possibility, for all $s>0$
fields there are only two physical degrees of freedom (or field polarizations) at space 
point. Therefore, in the dual description, each such a field is described by the 
spin system with two spins at point. However, the spin systems are characterized 
by very different types of interactions for each type of field. 

\section{Conclusions}

In this article we have extended discussion of \cite{Mielczarek:2016rax} by focusing 
on the case of spherical field space in the position representation, instead of the 
momentum representation. We have shown that the scalar field theory with spherical phase 
space can be considered as dual description of the continuous spin system. The concrete 
form of the scalar field theory depends on the particular spin interaction Hamiltonian. 
As an example, we have shown that the Heisenberg model coupled to an external magnetic
field corresponds (in the perturbative limit) to the non-relativistic scalar field theory with 
an additional $\frac{1}{2 M^2}(\nabla  \pi_\varphi)^2$ term in the Hamilton function. 

In the considered case we focused our attention on the leading order terms. However, relation between 
the field theories and the spin systems discussed here leads also to higher order interaction 
terms, which will require careful analysis in the future studies. In particular, the specific 
form of the interaction terms, which may also break Lorentz invariance, may turn out to 
be relevant from the viewpoint of renormalizability of the theory.  

On the basis of the scalar field example, we proposed the  \emph{Spin-Field correspondence} 
which extends the duality to the different types of fields considered in the context 
of fundamental interactions. The correspondence may play a technical role in constructing 
field theoretical description of condensed matter systems. Another application is more fundamental 
and provides a way to address the issue of the origin of existing types of fields. The correspondence 
suggests that all known types of fields can be considered as an emergent property of the 
spin systems. Furthermore, the correspondence provides geometric interpretation of the 
Hamiltonians considered in the field theory. In particular, the quadratic terms can be 
seen as the leading order contributions from such geometric objects as angles resulting 
from the scalar products of the type $\vec{J}_i \cdot \vec{J}_j$ and $\vec{J}_i \cdot \vec{B}$. 

In the case of the fundamental spin structure the question is, however, what the underlying 
spin system is? It might only be a coincidence, but the currently considered approaches 
to quantum gravity, such as Loop Quantum Gravity \cite{Ashtekar:2004eh} are heavily 
based on the mathematical structure of spin as well as spin networks and spin foams. 
One can hypothesize that the structure of this type (but perhaps more sophisticated) may 
play a role of discrete spin system behind the field theoretical description of matter and interactions 
(including gravity). In such a picture, different types of excitations of the underlying spin 
system would lead to effective field theoretical descriptions. Theoretical realization of this tempting 
possibility, will be a subject of our further studies. 

\section*{Acknowledgements}
This work is supported by the Iuventus Plus grant No.~0302/IP3/2015/73 from the 
Polish Ministry of Science and Higher Education. Author would like to thank to Tomasz 
Trze\'sniewski for his careful reading of the manuscript and helpful comments.

\end{document}